\documentclass[twocolumn,showpacs,superscriptaddress,showkeys,reprint,amsmath,amssymb,aps,prb]{revtex4-1}
\usepackage{graphicx,color}
\usepackage{dcolumn}
\usepackage{hyperref}
\usepackage{bm}
\usepackage{epstopdf}
\usepackage{amsmath}
\hfuzz=\maxdimen
\usepackage{float}

\usepackage{appendix}

\begin{document}

\title{Pairing symmetry in monolayer of orthorhombic CoSb}

\author{Tianzhong Yuan}
\affiliation{State Key Laboratory of Surface Physics and Department of Physics, Fudan University, Shanghai 200433, China}

\author{Muyuan Zou}
\affiliation{State Key Laboratory of Surface Physics and Department of Physics, Fudan University, Shanghai 200433, China}

\author{Wentao Jin}
\affiliation{Key Laboratory of Micro-Nano Measurement-Manipulation and Physics (Ministry of Education), School of Physics, Beihang University, Beijing 100191, China}

\author{Xinyuan Wei}
\affiliation{State Key Laboratory of Surface Physics and Department of Physics, Fudan University, Shanghai 200433, China}

\author{Xuguang Xu}
\email{xuxg@shanghaitech.edu.cn}
\affiliation{School of Physical Science and Technology, ShanghaiTech University, Shanghai 201210, China}

\author{Wei Li}
\email{w$_$li@fudan.edu.cn}
\affiliation{State Key Laboratory of Surface Physics and Department of Physics, Fudan University, Shanghai 200433, China}
\affiliation{Collaborative Innovation Center of Advanced Microstructures, Nanjing University, Jiangsu 210093, China}

\date{\today}

\begin{abstract}
Ferromagnetism and superconductivity are generally considered to be antagonistic phenomena in condensed matter physics. Here, we theoretically study the interplay between the ferromagnetic and superconducting orders in a recent discovered monolayered CoSb superconductor with an orthorhombic symmetry and net magnetization, and demonstrate the pairing symmetry of CoSb as a candidate of non-unitary superconductor with time-reversal symmetry breaking. By performing the group theory analysis and the first-principles calculations, the superconducting order parameter is suggested to be a triplet pairing with the irreducible representation of $^3B_{2u}$, which displays intriguing nodal points and non-zero periodic modulation of Cooper pair spin polarization on the Fermi surface topologies. These findings not only provide a significant theoretical insight into the coexistence of superconductivity and ferromagnetism, but also reveal the exotic spin polarized Cooper pairing driven by ferromagnetic spin fluctuations in a triplet superconductor.
\end{abstract}

\maketitle

\section{Introduction}

The search for exotic unconventional pairing superconductivity with time-reversal symmetry breaking is one of the most challenging tasks in condensed matter physics. Among them, the prominent chiral pairing superconductors originated from the contribution of orbital angular momentum of Cooper paired electrons. For example, the chiral $p$-wave pairing topological superconductors~\cite{Qi2011,Sato2017}, have received great attentions as they host the Majorana quasiparticles at the boundaries~\cite{Kitaev2001,Ivanov2001,LFu2008,KTLaw,Sau2010,GXu}, which is equivalent to the non-Abelian Moore-Read (Pfaffian) spin-triplet paired state in the fractional quantum Hall effect with filling factor of $5/2$~\cite{Green2000,Moore1991,Read1992}, and has potential applications in the topological quantum computing~\cite{Beenakker2013,Kitaev,Nayak2008,Alicea,Franz2015,Aguado}. Experimentally, the evidences of observing Majorana bound states have been extensively reported in various quantum systems, including the one-dimensional nanowires in contact with superconductors~\cite{Mourik2012,MDeng2012,ADas2012,Deng2016,HZhang2018}, at the edges of iron-atoms chains formed on the surface of superconducting lead~\cite{Perge2014}, at the interface between a topological insulator and an $s$-wave superconductor~\cite{JFJia2015,JFJia2016}, and the quantum spin liquids~\cite{Banerjee2016}, 
as well as the iron-based superconductors~\cite{DLFeng,JXYin2015,HHWen,Wang2018,SZhu2019,CChen2020,Wang2020}.

Additionally, another class of superconductors with time-reversal symmetry breaking, the intriguing non-unitary pairing superconductors~\cite{Sigrist}, originated from the contribution of spin angular momentum of Cooper paired electrons, are inspiring enormous research interests in the condensed matter communities recently. The richness of existing Majorana quasiparticles in three-dimensional high-symmetry non-unitary pairing superconductors has been theoretically proposed~\cite{Venderbos}. So far, however, the only experimentally established the non-unitary pairing is in the $A_1$ phase of superfluid $^3$He in an applied high magnetic field~\cite{Mermin,Leggett,Wheatley}, although the non-unitary paired states have been extensively reported in the heavy fermion superconductor UPt$_3$ related to the $B$ phase at low temperature in an applied magnetic field~\cite{Machida,Sauls,Tou,Joynt}, and in the noncentrosymmetric LaNiC$_2$~\cite{Cywinski,Quintanilla} and the centrosymmetric LaNiGa$_2$ superconductors~\cite{Hillier,Ghosh} with the absence of an applied magnetic field.

In this paper, we theoretically propose the monolayered orthorhombic CoSb as a candidate of non-unitary pairing superconductor, which has been successfully grown on the SrTiO$_3$(001) substrate by molecular beam epitaxy. Experimentally, symmetric superconducting gap around the Fermi level with coherence peaks at around $\pm$ 6 meV was observed by $in$-$situ$ scanning tunneling spectroscopy (STS), accompanied with a weak net ferromagnetic (FM) moment lying in the basal plane found by $ex$-$situ$ magnetization measurements~\cite{CDing2019}. The pairing symmetry of this system, however, remains elusive in experiments. Theoretically, the group symmetry analysis suggests the pairing symmetry of monolayered CoSb to be a non-unitary triplet gap function of $^3B_{2u}$ or $^3B_{3u}$ symmetry with nodes. Within the framework of density-functional theory, the calculations demonstrate the ground state of monolayered CoSb to be a half metal with the easy axis of FM magnetization along the $\hat{y}$ axis lying in the basal plane, which is consistent with experimental observation~\cite{CDing2019} and supports the group theory analysis that only spin-down electrons are responsible for the Cooper pairing in the non-unitary superconducting state. Using the strong-coupling approximation the superconducting order parameter in the monolayered CoSb is leaded to be a triplet pairing with the irreducible representation of $^3B_{2u}$, displaying intriguing nodal points and non-zero periodic modulation of Cooper pair spin polarization on the Fermi surface topologies. These findings imply the coexistence of ferromagnetism and superconductivity and the exotic spin polarized Cooper pairing driven by FM spin fluctuations in the superconductor CoSb.

\begin{figure*}
\centering{}\includegraphics[bb=10 10 580 150,width=16.5cm,height=3.8cm]{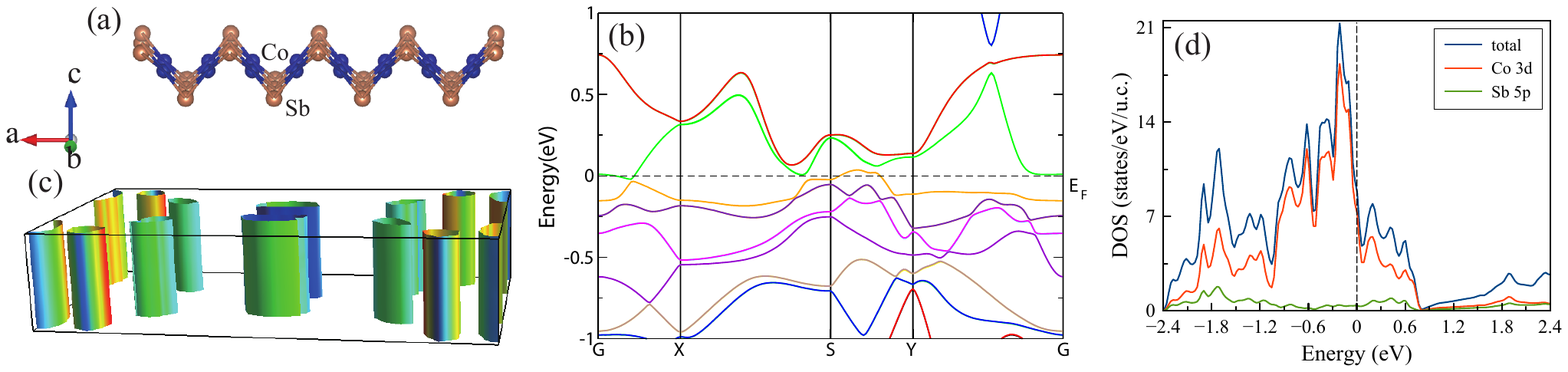}
\caption{(Color online) (a) The schematic illustration of the crystal structure of monolayered orthorhombic CoSb. (b) The electronic band structure and (c) the corresponding Fermi surface topologies for the NM state of monolayered CoSb. (d) The total DOS and PDOS on Co 3$d$ and Sb 5$p$ orbitals for the NM state of monolayered CoSb. The Fermi energies are set to zero.}
\label{fig1}
\end{figure*}

The rest of this paper is structured as follows. In section~\ref{Sec2}, we present the group symmetry analysis on the pairing symmetry of monolayered orthorhombic CoSb superconductor. The electronic and magnetic structures of monolayered CoSb are systemically addressed in section~\ref{Sec3} by using the first-principles calculations. The theoretical Bogoliubov-de Gennes (BdG) equation of superconductivity is clarified in section~\ref{Sec4}. Finally, we give a brief conclusion in section~\ref{Sec5}.

\section{Group symmetry analysis}\label{Sec2}

Considering the $D_{2h}$ point group of the superconducting orthorhombic CoSb monolayer shown in Fig.~\ref{fig1}(a) with the time-reversal symmetry breaking~\cite{CDing2019}, the superconducting pairing gap function $\Delta(\vec{k})$ can be factored into the basis functions with the irreducible representation of the group $SO(3)\times D_{2h}$ in the weak spin-orbit coupling limit~\cite{Annett,VPMineev,Samokhin}, where $\times$ and $SO(3)$ represent the direct product and all spin rotations, respectively. Similar to that in the centrosymmetric superconductor LaNiGa$_2$~\cite{Hillier}, this product group has a total of eight irreducible representations listed in Table~\ref{tabel1}, including four one-dimensional singlet representations and four three-dimensional triplet representations. The latter have a gap function that transforms like a vector under spin rotations, resulting in the two possible ground states in a general Ginzburg-Landau theory~\cite{Annett}. This leads to twelve possible gap functions listed in Table~\ref{tabel1}, of which eight are unitary and four are non-unitary. Only the four non-unitary gap functions are non-trivially complex that can break the time-reversal symmetry. Furthermore, in accordance with the two-dimensionality of monolayered CoSb, we further eliminate the two states showing strong $k_z$ dependence of the gap function. In contrast, there are only four possible gap functions and none of them could break time-reversal symmetry in the strong spin-orbit coupling limit, as listed in Table~\ref{tabel1}. Therefore, from the viewpoint of symmetry, the superconducting orthorhombic CoSb monolayer has to be a non-unitary triplet pairing superconductor with a weak spin-orbit coupling, so that the possible gap functions with $^3B_{2u}$ and $^3B_{3u}$ symmetry are only compatible with the experimental observation of time-reversal symmetry breaking~\cite{CDing2019}. In them, only spin-down electrons participate in pairing, and thus there is an ungapped Fermi surface coexisting with another one with nodes ($^3B_{2u}$ or $^3B_{3u}$).

\begin{table}[b!]
\caption{The upper and lower tables show the gap functions of the homogeneous superconducting states allowed by symmetry for a weak and a strong spin-orbit coupling, respectively. We have used the standard notation~\cite{Sigrist} $\hat{\Delta}(\vec{k})=\Delta(\vec{k})i\hat{\sigma}_y$ for singlet states and $\hat{\Delta}(\vec{k})=i[\mathbf{d}(\vec{k})\cdot\hat{\mathbf{\sigma}}]\hat{\sigma}_y$ for triplets, where $\mathbf{\hat{\sigma}}$ is the vector of Pauli matrices, and $\vec{k}$ is the momentum.}
\begin{ruledtabular} %
\begin{tabular}{ccc}
$SO(3)\times D_{2h}$  & unitary state & non-unitary state   \\
\hline
$^1A_{1g}$ & $\Delta(\vec{k})=1$& 0 \\
\hline
$^1B_{1g}$ & $\Delta(\vec{k})=k_xk_y$& 0 \\
\hline
$^1B_{2g}$ & $\Delta(\vec{k})=k_xk_z$& 0 \\
\hline
$^1B_{3g}$ & $\Delta(\vec{k})=k_yk_z$& 0 \\
\hline
$^3A_{1u}$ & $d(\vec{k})=(0,0,1)k_xk_yk_z$& $d(\vec{k})=(1,-i,0)k_xk_yk_z$ \\
\hline
$^3B_{1u}$ & $d(\vec{k})=(0,0,1)k_z$& $d(\vec{k})=(1,-i,0)k_z$ \\
\hline
$^3B_{2u}$ & $d(\vec{k})=(0,0,1)k_y$& $d(\vec{k})=(1,-i,0)k_y$ \\
\hline
$^3B_{3u}$ & $d(\vec{k})=(0,0,1)k_x$& $d(\vec{k})=(1,-i,0)k_x$ \\
\end{tabular}
\begin{tabular}{cc}
$D_{2h}$  & Gap functions with strong spin-orbit coupling   \\
\hline
$A_{1u}$ & $d(\vec{k})=(Ak_x, Bk_y, Ck_z)$ \\
\hline
$B_{1u}$ & $d(\vec{k})=(Ak_y, Bk_x, Ck_xk_yk_z)$ \\
\hline
$B_{2u}$ & $d(\vec{k})=(Ak_z, Bk_xk_yk_z, Ck_x)$ \\
\hline
$B_{3u}$ & $d(\vec{k})=(Ak_xk_yk_z, Bk_z, Ck_y)$ \\
\end{tabular}
\end{ruledtabular}
\label{tabel1}
\end{table}

\begin{figure*}
\centering{}\includegraphics[bb=10 10 530 135,width=16.5cm,height=3.8cm]{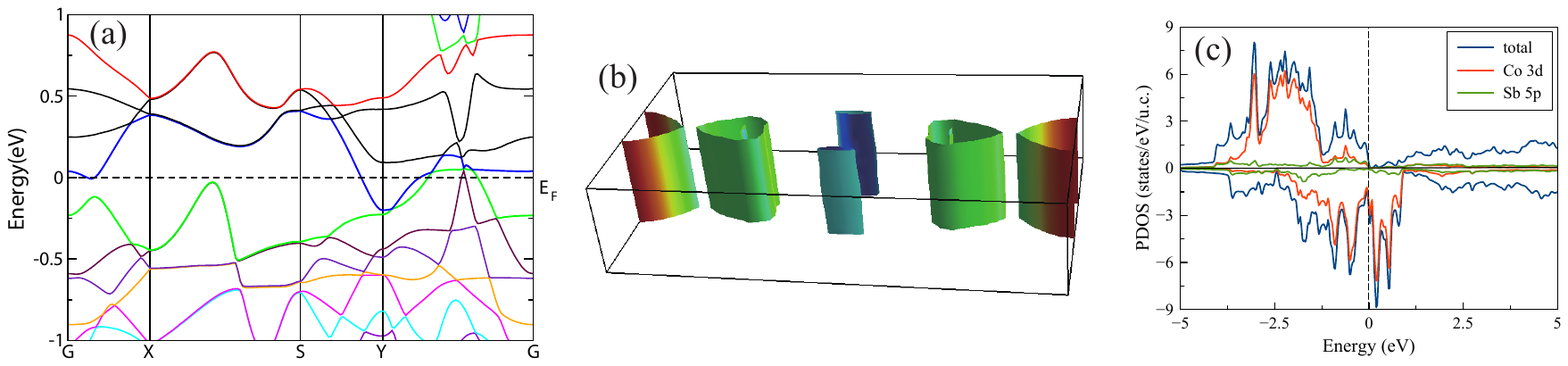}
\caption{(Color online) (a) The electronic band structure and (b) the corresponding Fermi surface topologies for the FM state of monolayered CoSb. (c) The PDOS on the spin-up and spin-down species of total, Co 3$d$, and Sb 5$p$ orbitals for the FM state of monolayered CoSb. The Fermi energies are set to zero.}
\label{fig2}
\end{figure*}

\section{The First-Principles Calculations}\label{Sec3}
The first-principles calculations are performed using the all-electron full potential linear augmented plane wave method~\cite{Singh} as implemented in the WIEN2k code~\cite{wien2k}. The exchange-correlation potential is calculated using the generalized gradient approximation as proposed by Perdew, Burke, and Ernzerhof~\cite{PBE1996}. Although the conduction electrons mainly originated from the light atoms of cobalt have a weak spin-orbit coupling, consistent with the group analysis, the heavy mediated anion of antimony has a strong spin-orbit coupling, whose strength is proportional to $Z^4$ (where $Z$ is the atomic number; $Z=51$ for Sb)~\cite{Li2014}, leading to a significant changes of the overlapped wave functions between the Co 3$d$ and Sb 5$p$ orbitals~\cite{Kanamori}. Therefore, the spin-orbit coupling is included with the second variational method throughout the calculations. Furthermore, a 3000 $\vec{k}$-point is chosen to ensure the calculation with an accuracy of $10^{-5}$ eV, and all structural parameters (lattice constants, $a_1=5.92$ \AA\ and $a_2=3.24$ \AA, as well as internal coordinates) are performed using the values of experimental crystal structure~\cite{CDing2019} shown in Fig.~\ref{fig1}(a). To reduce the interaction between neighboring layers of CoSb, a vacuum slab of 15 \AA\ along the $\hat{z}$ axis is introduced.

Figs.~\ref{fig1}(b)-(d) show the non-magnetic (NM) electronic structures of monolayered orthorhombic CoSb, where no spin polarization is allowed on the Co ions. Such a study can provide a benchmark for inspecting whether the magnetically ordered state is favorable. From the calculated energy band structure and the corresponding Fermi surface topologies shown in Figs.~\ref{fig1}(b) and (c), there are mainly two bands crossing the Fermi level contributing to the electron conduction in orthorhombic CoSb, in contrast to the four bands across the Fermi level in the tetragonal CoSb~\cite{Zhang2020,myzou}. Verifying the orbital characters of the energy bands around the Fermi level (see details in Fig.~\ref{fig_S1} in~\ref{SecA2}), we notice that the five Co 3$d$ orbitals participate the electron conductions, implying the strong Hund's coupling in the Co 3$d$ orbitals.

The calculated density of states (DOS) and the projected DOS (PDOS) on Co $3d$ and Sb $5p$ orbitals for the NM state of monolayered orthorhombic CoSb are shown in Fig.~\ref{fig1}(d). It can be seen that the conduction electrons mainly come from the contribution of Co $3d$ states partially hybridized with mediated Sb $5p$ states. Inspecting the value of DOS at the Fermi level, $N(E_f)=3.58$ states per eV per Co atom, we notice that this value is much larger than that in the tetragonal CoSb~\cite{Zhang2020,myzou} and the iron-based superconductors~\cite{WLi2012,WLiFOP2015}. While magnetism may occur with lower values of the DOS, it must occur within a band picture if the Stoner criterion~\cite{WLi2012,Singh2008,WLiFOP2018}, $N(E_f)\times I > 1$, is met, where $I$ is the Stoner parameter, taking values of $0.7-0.9$ eV for ions near the middle of the $3d$ series (note that the effective $I$ can be reduced by hybridization)~\cite{Singh2008}, implying the NM state is unstable against the magnetic states for monolayered CoSb.

\begin{figure*}
\centering{}\includegraphics[bb=30 55 545 280,width=16cm,height=6.5cm]{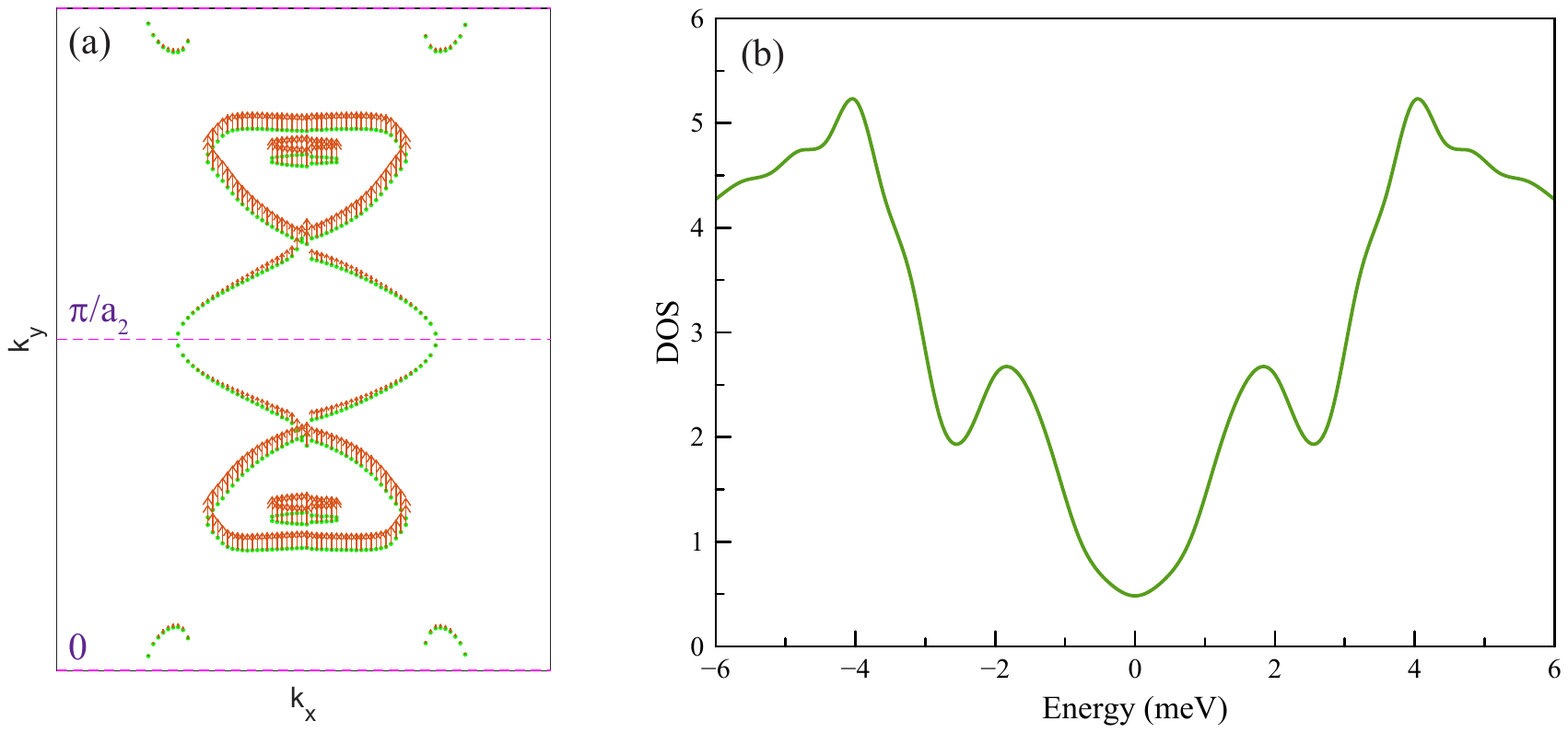}
\caption{(Color online) (a) A schematic plot for the gap nodal structure of non-unitary pairing on the Fermi surface topologies. The green dotted lines denote the Fermi surface topology of monolayered CoSb, and the magenta dashed lines denote the zero gap value of the order parameter $d(\vec{k})=(1,-i, 0)\sin(k_ya_2)$ with the irreducible representation of $^3B_{2u}$. The vector plot of Cooper pair spin polarization $\langle \hat{S}_{\vec{k}}\rangle$ is also shown on the counters of Fermi surface topologies. (b) The DOS as a function of energy for the non-unitary superconducting state. The parameter of pairing amplitude is set as $\Delta_0=5$ meV.}
\label{fig3}
\end{figure*}

In order to capture the magnetic behavior of Co 3$d$ states in the monolayered orthorhombic CoSb, we consider a two-dimensional phenomenologically theoretical Heisenberg model on the Co ion sites as follow~\cite{myzou,WLi2012}:
\begin{equation}
\hat{H} = J_{1x}\sum_i\vec{S}_i\vec{S}_{i+\hat{x}}+J_{1y}\sum_i\vec{S}_i\vec{S}_{i+\hat{y}}+J_2\sum_{\langle\langle i,j\rangle\rangle}\vec{S}_i\vec{S}_j,
\label{Eq1}
\end{equation}
where $\vec{S}$ is the magnitude of Co spin. The $\langle\langle i,j\rangle\rangle$ denotes the summation over the next-nearest neighbor Co ion sites. The parameters $J_{1x}$ and $J_{1y}$ describe the nearest neighboring exchange interactions along the $\hat{x}$ and $\hat{y}$ direction shown in Fig.~\ref{fig1}(a) with the labels of $a$ and $b$, respectively, and $J_{2}$ denotes the next-nearest neighboring exchange interaction. From the calculated energies for various magnetic configurations~\cite{ZZhou2019} (see~\ref{SecA1} in details), the magnetic exchange couplings $J_{1x}=1.05$ meV, $J_{1y} =-46.46$ meV, and $J_{2}=-2.98$ meV are found for the monolayered orthorhombic CoSb. The strong FM superexchange coupling strength along the $\hat{y}$ axis could be understood through the Goodenough-Kanamori orthogonal rule~\cite{Kanamori,Goodenough} at first glance that the interacting cations of Co atoms connected to the intervening anions of Sb form an angle of $74.1^{\circ} (\sim 90^{\circ})$, which promotes the mediated Sb 5$p$ orbitals to be orthogonal to the two nearest neighboring Co 3$d$ orbitals. Considering the strong Hund's coupling on Co 3$d$ orbitals, the Co$^{3+}$ ion with 3$d^{6}$ electronic configuration favors the unpaired spin on Sb 5$p$ orbitals to be parallelly aligned to the spin of the Co 3$d$ orbitals, resulting in the FM exchange coupling. However, when the distance of $3.24$ \AA\ ($=a_2$) between two nearest Co ion sites along $\hat{y}$ axis is changed to $2.96$ \AA\ ($=a_1/2$) along the $\hat{x}$ axis and the bond angle of Co-Sb-Co is changed to $66.8^{\circ}$, the orthogonality between the Sb 5$p$ and the two nearest neighboring Co 3$d$ orbitals is weakened significantly and thus the antiferromagnetic superexchange coupling could be gradually emerged along the $\hat{x}$ direction. Due to the strong FM exchange couplings on the CoSb layer, it suggests the ground state of CoSb to be a FM order~\cite{Maekawa2004}, which is consistent with the magnetization measurements on the monolayered films of orthorhombic CoSb~\cite{CDing2019}. Furthermore, the strong anisotropic FM superexchange interaction along the $\hat{y}$ axis drives the easy axis of magnetization of CoSb towards the $\hat{y}$ axis lying in the basal plane, which is also confirmed by the total energy calculations. The magnetic momentum of 1.83 $\mu_B$ on Co ion sites are found (see details in Table~\ref{tabelS1} in~\ref{SecA1}).

The calculated low-energy band structure, the corresponding Fermi surface topologies, and the PDOS on the spin-up and spin-down species of total, Co $3d$ and Sb $5p$ orbitals for the FM ordered state with fixed magnetization along the $\hat{y}$ axis in the monolayered orthorhombic CoSb are shown in Fig.~\ref{fig2}. Compared with the NM state shown in Fig.~\ref{fig1}, we find that most of the bands around the Fermi level are gapped by the FM order. The corresponding electronic DOS at the Fermi level is $N(E_f)=0.12$ and $N(E_f)=1.48$ states per eV per Co atom for spin-up and spin-down species, respectively, which are significantly less than that of the NM state (3.58 states per eV per Co atom), demonstrating a half metal nature of the monolayered orthorhombic CoSb that the spin-up orbitals are fully occupied while the spin-down orbitals are partially occupied (see details in Figs.~\ref{fig_S2} and~\ref{fig_S3} in~\ref{SecA2}). This finding is consistent with the group theory analysis that only spin-down electrons are responsible for the Cooper pairs in the non-unitary superconducting state.

\section{Theoretical Model Calculations}\label{Sec4}

A simplified theoretical model of low-energy excitations in the non-unitary superconducting state is provided for further understanding of the behaviors of superconducting electrons based on the following BdG Hamiltonian:
\begin{equation}
\hat{H}_{sc} = \left(
                 \begin{array}{cc}
                   \hat{H}_0(\vec{k}) & \hat{\Delta}_0(\vec{k}) \\
                   \hat{\Delta}_0^{\dag}(\vec{k}) &-\hat{H}_0(\vec{k}) \\
                 \end{array}
               \right),
\label{Eq2}
\end{equation}
where $\vec{k}$ is the momentum of the excitation, $\hat{H}_0(\vec{k})$ describes an effective spin-dependent four-band normal-state free electron Hamiltonian obtained through an interpolation method by projecting the first-principles calculated bands shown in Fig.~\ref{fig2} onto the lowest two spin-dependent bands around the Fermi level~\cite{WLiPairing,FTS} in the momentum space, and $\hat{\Delta}_0(\vec{k})=\Delta_0\hat{\Delta}(\vec{k})\otimes i\tau_y$ represents the pairing potential with a pairing amplitude of $\Delta_0$. In the tensor products, the first sector represents the spin channels $\sigma=\uparrow, \downarrow$ shown in the caption of Table~\ref{tabel1} while the second represents the two band channels~\cite{Ghosh}. Following the group symmetry analysis, the $\mathbf{d}(\vec{k})$ vector has two possible choices of $d(\vec{k})=(1,-i, 0)\sin(k_ya_2)$ and $d(\vec{k})=(1,-i, 0)\sin(k_xa_1)$, as listed in the Table~\ref{tabel1}, corresponding to the irreducible representations of $^3B_{2u}$ and $^3B_{3u}$, respectively. Here we have assumed that the Cooper pairs carry the spin magnetization with the value of $\langle \hat{S}_{\vec{k}}\rangle=i\mathbf{d}\times\mathbf{d}^*$~\cite{Sigrist} along the $\hat{y}$ axis in accordance with the FM magnetization obtained by the first-principles calculations. Since the pairing amplitude of $\Delta_0$ is proportional to FM superexchange coupling strength within strong-coupling approximation, the triplet pairing state with the irreducible representations of $^3B_{3u}$ is energetically unfavorable rather than that of $^3B_{2u}$, to avoid the short-range repulsion caused by the antiferromagnetic superexchange coupling along the $\hat{x}$ axis~\cite{WLiPairing,Tsunetsugu,Fujimoto,Kallin}, which can also been seen clearly in~\ref{SecA3}. Therefore, the non-unitary paired $^3B_{2u}$ state induced by the FM spin fluctuations results in the formation of Cooper pairing in monolayered CoSb superconductor. The gap zeros of $^3B_{2u}$ state ($k_y=0$ and $k_y=\pi/a_2$) cross the Fermi surface topologies, shown in Fig.~\ref{fig3}(a), leading to intriguing nodal behavior. Additionally, it is interesting to point out that the amplitude of Cooper pair spin polarization $\langle \hat{S}_{\vec{k}}\rangle$ on the counters of Fermi surface topologies displays a periodic modulations and vanishes at the nodal points on the Fermi surface topologies, which are the typical characters of non-unitary pairing superconductivity. The DOS of superconducting state with the non-unitary pairing of $^3B_{2u}$ symmetry is also calculated and shown in Fig.~\ref{fig3}(b). As is expected, the V-shaped DOS is clearly visible, qualitatively consistent with the experimentally observed STS spectra~\cite{CDing2019}.


\section{Conclusion}\label{Sec5}
By performing the group theory analysis and the first-principles calculations, we systemically study the electronic and magnetic properties in the monolayered orthorhombic CoSb superconductor, and find the normal state of CoSb to be a half metal with the easy axis of FM magnetization along the $\hat{y}$ axis lying in the basal plane, suggesting the orthorhombic CoSb as a candidate of non-unitary superconductor in which only spin-down electrons are responsible for the Cooper pairing. Using the strong-coupling approximation, we suggest the pairing symmetry of CoSb to be a triplet pairing with the irreducible representations of $^3B_{2u}$ that displays intriguing nodal points and non-zero periodic modulation of Cooper pair spin polarization on the Fermi surface topologies. These findings imply the novel coexistence of FM and superconducting orders in CoSb and the exotic spin polarized Cooper pairing driven by FM spin fluctuations in a triplet superconductor.

\section*{ACKNOWLEDGMENTS}

This work was supported by the National Natural Science Foundation of China (Grant No. 11927807) and the Natural Science Foundation of Shanghai of China (Grant Nos. 19ZR1402600 and 20DZ1100604). W. L. also acknowledges the start-up funding from Fudan University.

\begin{appendix}
\setcounter{figure}{0}
\renewcommand{\thefigure}{A\arabic{figure}}

\setcounter{table}{0}
\renewcommand{\thetable}{A\arabic{table}}

\section{Theoretical calculations with various magnetic configurations and the orientations dependent magnetic state calculations}\label{SecA1}

To explore the ground state with magnetic ordering on CoSb monolayer and evaluate the values of magnetic exchange coupling constants in Heisenberg model in Hamiltonian (\ref{Eq1}) in the main text, we perform the calculations with four possible magnetically ordered states in the Co layer with FM, AFM-1 (ferromagnetically along the $\hat{y}$-axis and antiferromagnetically along the $\hat{x}$-axis), AFM-2 (ferromagnetically along the $\hat{x}$-axis and antiferromagnetically along the $\hat{y}$-axis), and the N\'{e}el-AFM orders. The corresponding total energies of various magnetic ordering states are listed in Table~\ref{tabelA1}. It is clearly shown that the FM ordering state has the lowest total energy on CoSb monolayer. The calculated corresponding magnetic moment is 1.84 $\mu_B$ on Co atoms, which mainly originates from the contributions of localized electrons on Co sites. Furthermore, through these calculated energies the magnetic exchange coupling constants are also evaluated to be $J_{1x}=1.05$ meV, $J_{1y} =-46.46$ meV, and $J_{2}=-2.98$ meV. Additionally, we also perform the calculations on the energetic properties of the various Co spin ordered orientations for monolayered orthorhombic CoSb with FM ordering state, listed in Table~\ref{tabelS1}. The calculations demonstrate the ground state of monolayered CoSb is a ferromagnetic order with the magnetization along the $\hat{y}$ axis lying in the basal plane and the corresponding magnetic momentum of 1.83 $\mu_B$ on Co ion sites. Such magnetic anisotropy mainly originates from the single-ion contribution~\cite{Maekawa2004}.

\begin{table}
\caption{The calculated total energy of various magnetically ordered states of CoSb monolayer. $\Delta E$ is the total energy difference per Co atom with respect to the NM state, and $m_{Co}$ is the local magnetic moment per Co atom.}
\begin{ruledtabular} %
\begin{tabular}{ccccc}
CoSb  & FM & AFM-1 & AFM-2 & N\'{e}el-AFM   \\
\hline
$\Delta E$ (meV/Co)&$\mathbf{-43.48}$ &-42.04 &-28.14& -30.13 \\
\hline
$m_{Co}$($\mu_{B}$)& $\mathbf{1.84}$  & 1.73 & 1.67  & 1.66 \\
\end{tabular}\end{ruledtabular}
\label{tabelA1}
\end{table}

\begin{table}
\caption{Energetic properties of the different Co spin configurations for monolayered orthorhombic CoSb. Results are the total energy difference per Co atom for different Co spin directions in the ferromagnetic CoSb layer.}
\begin{ruledtabular} %
\begin{tabular}{cccc}
CoSb  & (100) & (010) & (001)   \\
\hline
$\Delta E$ (meV/Co)&0.13 & $\mathbf{0.0}$ &0.03 \\
\hline
$m_{Co}$($\mu_{B}$)& 1.83  & $\mathbf{1.83}$& 1.84   \\
\end{tabular}\end{ruledtabular}
\label{tabelS1}
\end{table}

\section{Orbital resolved band structure calculations}\label{SecA2}

\begin{figure*}
\centering{}\includegraphics[bb=10 10 505 220,width=17cm,height=6.8cm]{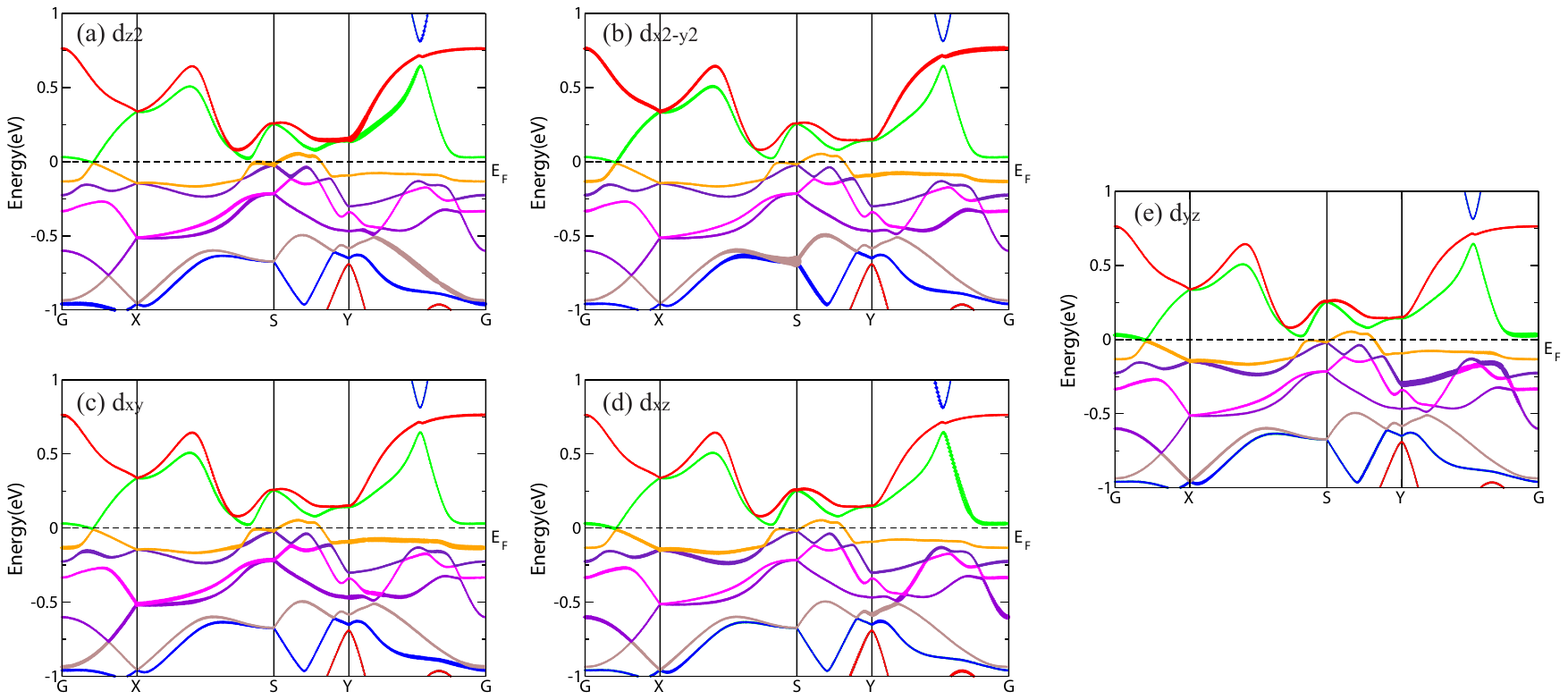}
\caption{(Color online) The orbitals resolved energy bands projected onto the Co $3d$ for the NM state of monolayered orthorhombic CoSb. The Fermi energies are set to zero.}
\label{fig_S1}
\end{figure*}

\begin{figure*}
\centering{}\includegraphics[bb=10 10 430 755,width=15cm,height=22cm]{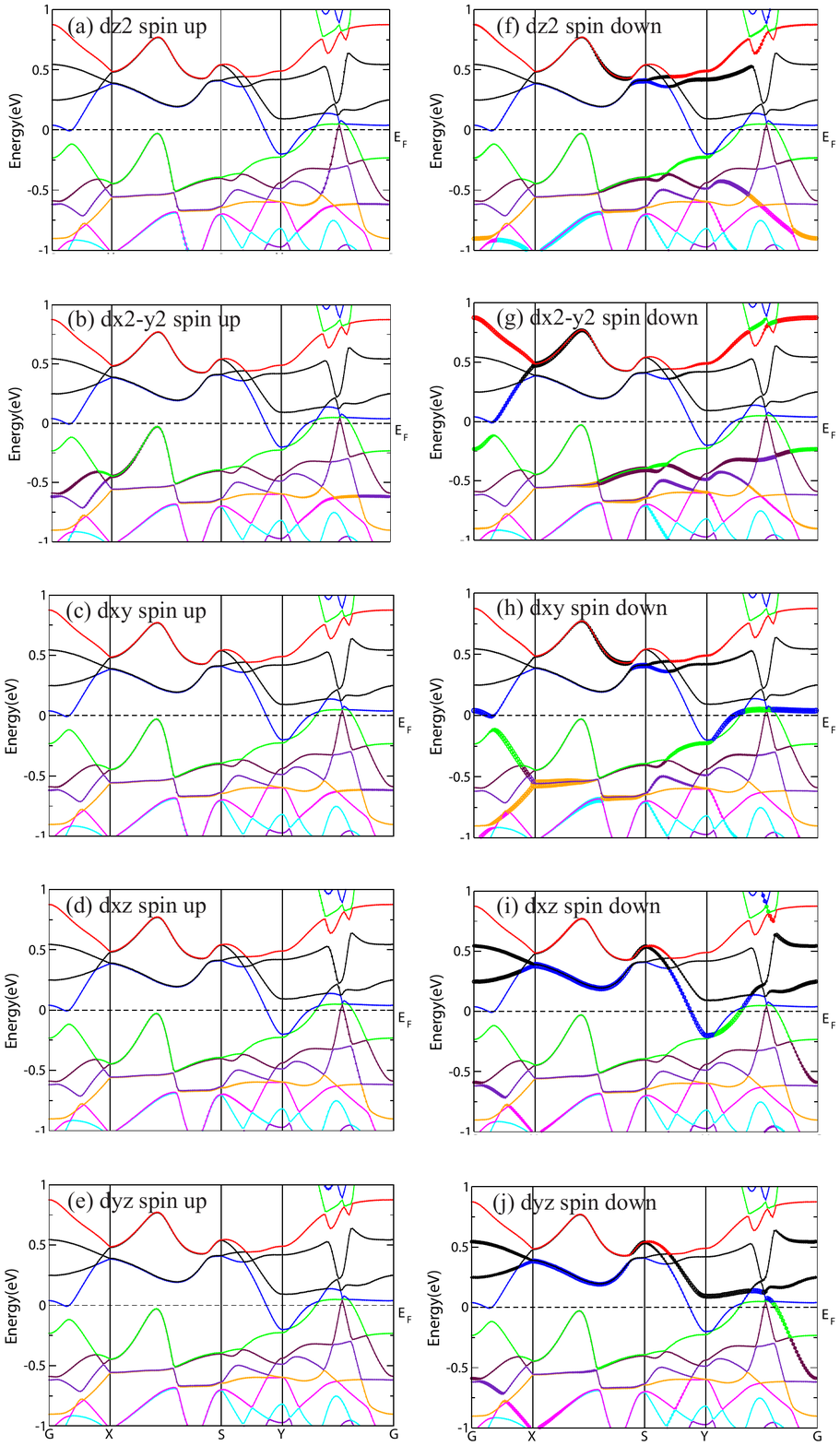}
\caption{(Color online) The spin dependent orbitals resolved energy bands projected by the Co $3d$ for the FM state of monolayered orthorhombic CoSb with the magnetization along the $\hat{y}$ axis lying in the CoSb plane. The Fermi energies are set to zero.}
\label{fig_S2}
\end{figure*}

\begin{figure*}
\centering{}\includegraphics[bb=10 10 480 480,width=15cm,height=16cm]{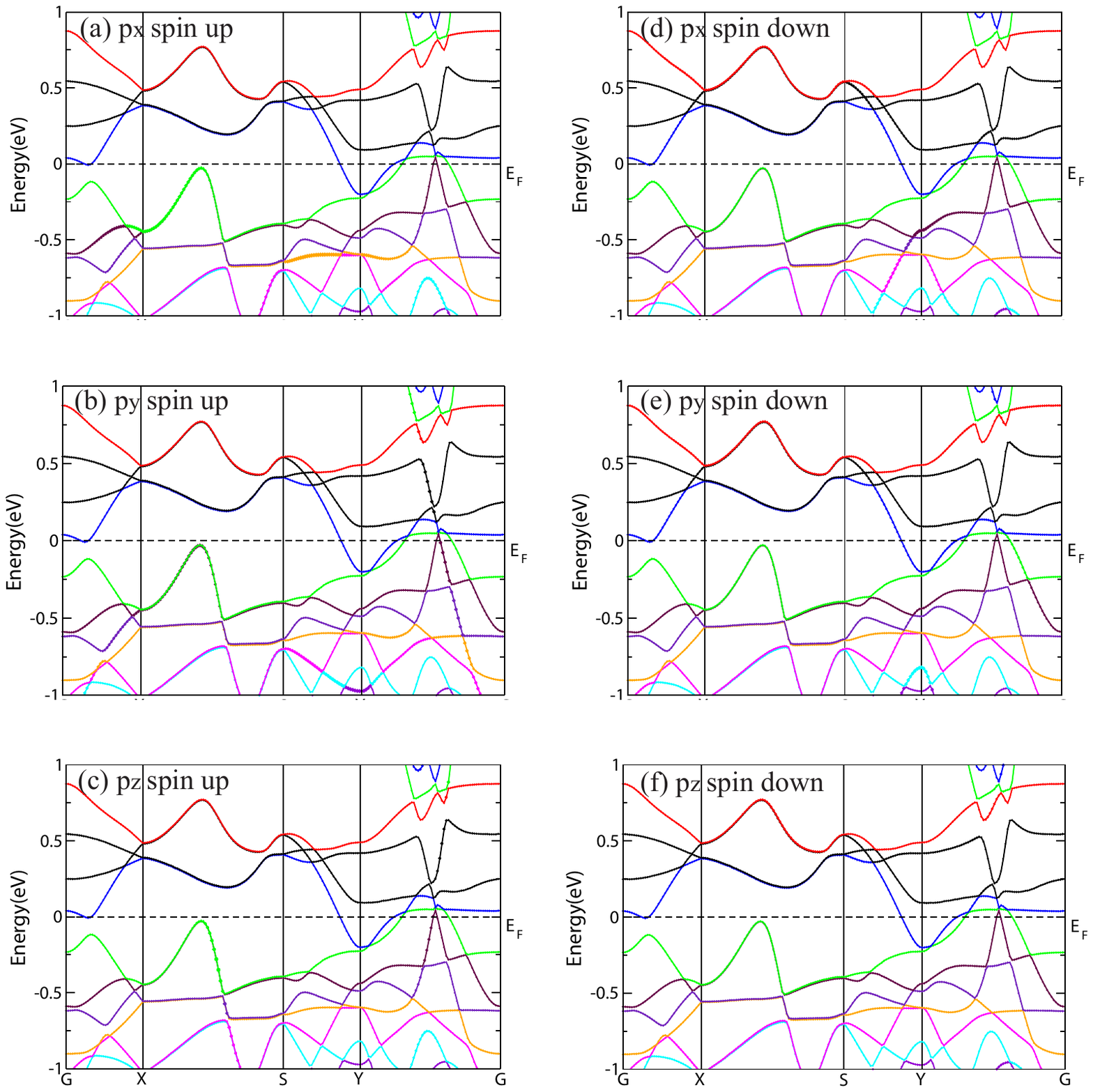}
\caption{(Color online) The spin dependent orbitals resolved energy bands projected by the Sb $5d$ for the FM state of monolayered orthorhombic CoSb with the magnetization along the $\hat{y}$ axis lying in the CoSb plane. The Fermi energies are set to zero.}
\label{fig_S3}
\end{figure*}

In this appendix, we present the detailed calculations of the orbital resolved energy bands of the NM and FM states of monolayered orthorhombic CoSb, as shown in Fig.~\ref{fig_S1}-~\ref{fig_S3}. Here it is interesting to point out that a Dirac like pocket appeared at the high symmetric line of $Y-G$ shown in Fig.~\ref{fig2} in the main text mainly stems from the contributions of spin-up components of Sb 5$p$ orbitals by inspecting the orbital resolved energy bands shown in Fig.~\ref{fig_S3}.

\section{Mean-field level of superconducting pairing}\label{SecA3}

Using the mean-field method of strong coupling limit for the superconducting CoSb monolayer, the superconducting pairing is decoupled from the magnetic Heisenberg model in Hamiltonian (\ref{Eq1}),
\begin{eqnarray}
\hat{H}^{pairing} &=& J_{1x}\sum_i(\Delta^{\uparrow\uparrow}_{1x}\hat{c}^{\dag}_{i,\uparrow}\hat{c}^{\dag}_{i+\hat{x},\uparrow}
+\Delta^{\downarrow\downarrow}_{1x}\hat{c}^{\dag}_{i,\downarrow}\hat{c}^{\dag}_{i+\hat{x},\downarrow}\nonumber\\
&-&\Delta^{\uparrow\downarrow}_{1x}\hat{c}^{\dag}_{i,\uparrow}\hat{c}^{\dag}_{i+\hat{x},\downarrow}
-\Delta^{\downarrow\uparrow}_{1x}\hat{c}^{\dag}_{i,\downarrow}\hat{c}^{\dag}_{i+\hat{x},\uparrow}) \nonumber\\
&+& J_{1y}\sum_i(\Delta^{\uparrow\uparrow}_{1y}\hat{c}^{\dag}_{i,\uparrow}\hat{c}^{\dag}_{i+\hat{y},\uparrow}
+\Delta^{\downarrow\downarrow}_{1y}\hat{c}^{\dag}_{i,\downarrow}\hat{c}^{\dag}_{i+\hat{y},\downarrow}\nonumber\\
&-&\Delta^{\uparrow\downarrow}_{1y}\hat{c}^{\dag}_{i,\uparrow}\hat{c}^{\dag}_{i+\hat{y},\downarrow}
-\Delta^{\downarrow\uparrow}_{1y}\hat{c}^{\dag}_{i,\downarrow}\hat{c}^{\dag}_{i+\hat{y},\uparrow})\nonumber\\
&+& J_2\sum_{\langle\langle i,j\rangle\rangle}(\Delta^{\uparrow\uparrow}_{ij}\hat{c}^{\dag}_{i,\uparrow}\hat{c}^{\dag}_{j,\uparrow}
+\Delta^{\downarrow\downarrow}_{ij}\hat{c}^{\dag}_{i,\downarrow}\hat{c}^{\dag}_{j,\downarrow}\nonumber\\
&-&\Delta^{\uparrow\downarrow}_{ij}\hat{c}^{\dag}_{i,\uparrow}\hat{c}^{\dag}_{j,\downarrow}
-\Delta^{\downarrow\uparrow}_{ij}\hat{c}^{\dag}_{i,\downarrow}\hat{c}^{\dag}_{j,\uparrow}) + h.c.,
\label{Eq3}
\end{eqnarray}
where $\Delta_{ij}^{\alpha\beta} (\alpha,\beta=\uparrow, \downarrow)$ are the superconducting pairing order parameters resulted from the magnetic spin fluctuations. Considering that the calculated magnitude of magnetic exchange coupling strength of $J_{1y}$ is significantly much larger than that of $J_{1x}$ and $J_{2}$, the pairing amplitude along the $\hat{y}$-axis, $\Delta_{1y}^{\alpha\beta}$, dominates in superconductivity. Furthermore, since the $J_{1y}$ is negative, the pairing orders with parallel spin orientation, $\Delta_{1y}^{\uparrow\uparrow}$ and $\Delta_{1y}^{\downarrow\downarrow}$, are energetically favorable rather than that with antiparallel ones. Therefore, the superconducting pairing order for CoSb superconductor has a typical spin-triplet, resulting from the ferromagnetic spin fluctuations~\cite{myzou}. Further performing the Fourier transformation, the pairing order parameters of $\Delta_{1y}^{\uparrow\uparrow}$ and $\Delta_{1y}^{\downarrow\downarrow}$ are proportional to a formation of $\sin(k_ya_2)$, where $a_2$ is the lattice constant of CoSb monolayer along the $\hat{y}$-axis shown in Fig.~\ref{fig1}(a) in the main text, in accordance with the group theory analytic result of pairing symmetry $^3B_{2u}$.

\end{appendix}

\end{document}